\documentclass[twocolumn,floatfix,showpacs]{revtex4}
\usepackage{graphicx}
\usepackage{amssymb}
\begin{document}

\title{Aharonov-Bohm effect for a valley-polarized current in graphene}
\author{A. Rycerz}
\affiliation{Marian Smoluchowski Institute of Physics, Jagiellonian University, Reymonta 4, 30-059 Krak\'{o}w, Poland}
\author{C. W. J. Beenakker}
\affiliation{Instituut-Lorentz, Universiteit Leiden, P.O. Box 9506, 2300 RA Leiden, The Netherlands}

\begin{abstract}
This is a numerical study of the conductance of an Aharonov-Bohm interferometer in a tight-binding model of graphene. Two single-mode ballistic point contacts with zigzag edges are connected by two arms of a hexagonal ring enclosing a magnetic flux $\Phi$. The point contacts function as valley filters, transmitting electrons from one valley of the band structure and reflecting electrons from the other valley. We find, in the wider rings, that the magnetoconductance oscillations with the fundamental periodicity $\Delta\Phi=h/e$ are suppressed when the two valley filters have opposite polarity, while the second and higher harmonics are unaffected or enhanced. This frequency doubling is interpreted in terms of a larger probability of intervalley scattering for electrons that travel several times around the ring. In the narrowest rings the current is blocked for any polarity of the valley filters, with small, nearly sinusoidal magnetoconductance oscillations. Qualitatively similar results are obtained if the hexagonal ring is replaced by a ring with an irregular boundary.
\end{abstract}

\date{September, 2007}
\maketitle

\section{Introduction}
The isolation of single two-dimensional layers of carbon (graphene) \cite{Gei07} has led experimental and theoretical physicists to reexamine classic effects from mesoscopic physics \cite{Imr96}. The search is for novel features that arise from the unusual conical band structure of a carbon monolayer \cite{Wal47}. One effect which has so far received little attention is the periodic magnetoconductance oscillation in a ring known as the Aharonov-Bohm effect \cite{Aha59,Imr89}. Preliminary experiments on the magnetoconductance of graphene rings have been reported by several groups \cite{Oez07,Mor07}. Theoretically, the energy spectrum of a {\em closed\/} graphene ring was studied as a function of the enclosed magnetic flux $\Phi$ by Recher et al.\ \cite{Rec07}.  

Here we study by computer simulation the electron transport through an {\em open\/} graphene ring, contacted to electron reservoirs by ballistic point contacts. Our work builds on an earlier finding \cite{Ryc07} that a single-mode point contact with zigzag edges operates as a valley filter. Depending on whether the Fermi level in the point contact lies in the conduction or valence band, the transmitted electrons occupy states in one or the other valley of the band structure. It was also shown in Ref.\ \onlinecite{Ryc07} that two adjacent valley filters function as a highly effective valley valve, passing or blocking the current depending on whether the two filters have the same or opposite polarity.

We find that the magnetoconductance oscillations in a multimode graphene ring with valley-filtering point contacts show the expected $\Delta\Phi=h/e$ periodicity if the two filters have the same polarity. For opposite polarity, however, a period doubling appears: The lowest harmonic is suppressed while the second and higher harmonics are unaffected or enhanced. We attribute the period doubling to intervalley scattering, which is more effective for electrons that have travelled more than once along the ring.

In few-mode rings the period doubling does not happen, instead the conductance is strongly suppressed with small, nearly sinusoidal, magnetoconductance oscillations. They appear in their clearest form in a hexagonal ring, but we find qualitatively the same behavior in a more generic ring with irregular boundaries.

\section{Aharonov-Bohm interferometer in graphene}

\begin{figure}[!ht]
\centerline{\includegraphics[width=\linewidth]{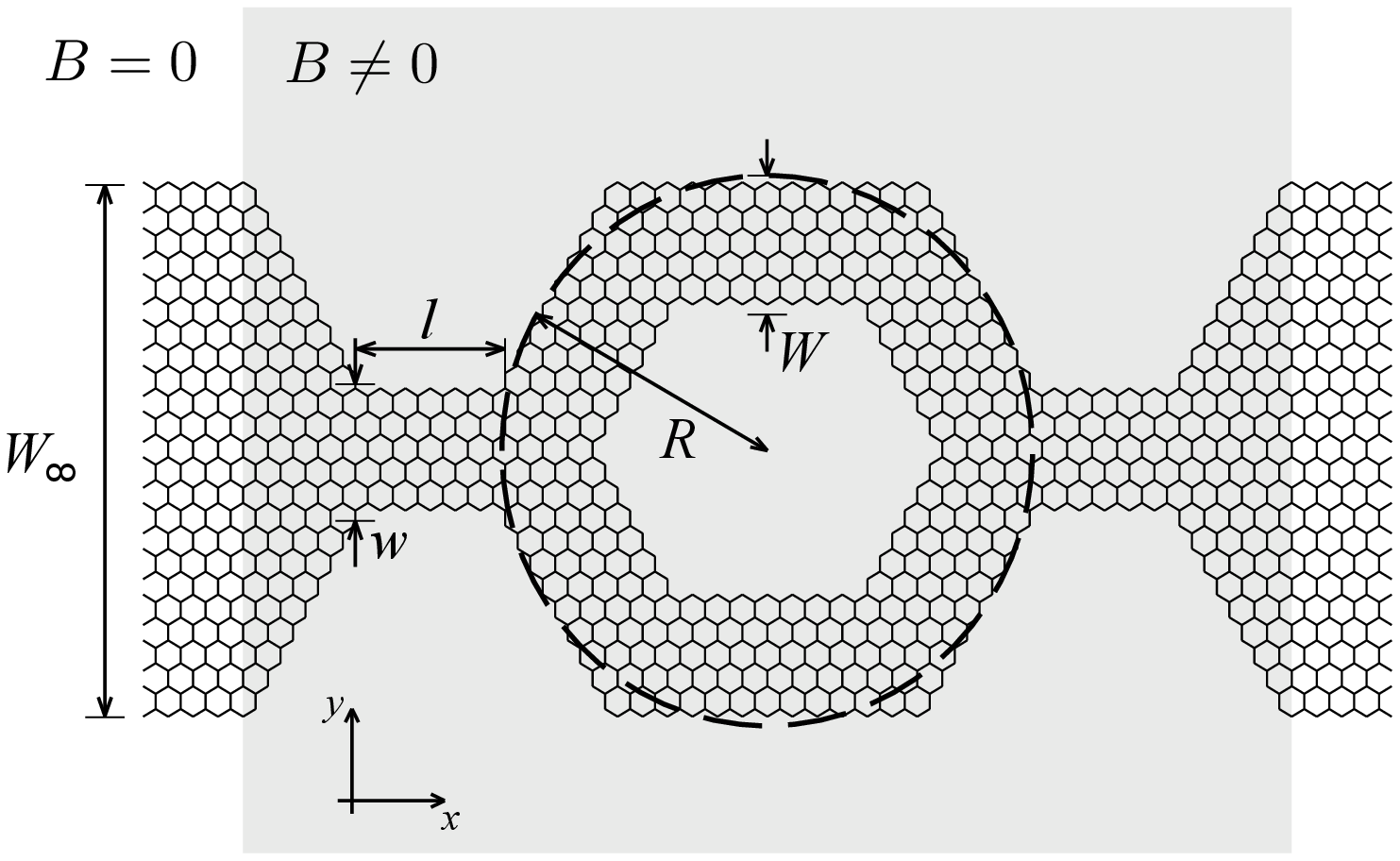}}

\centerline{\includegraphics[width=\linewidth]{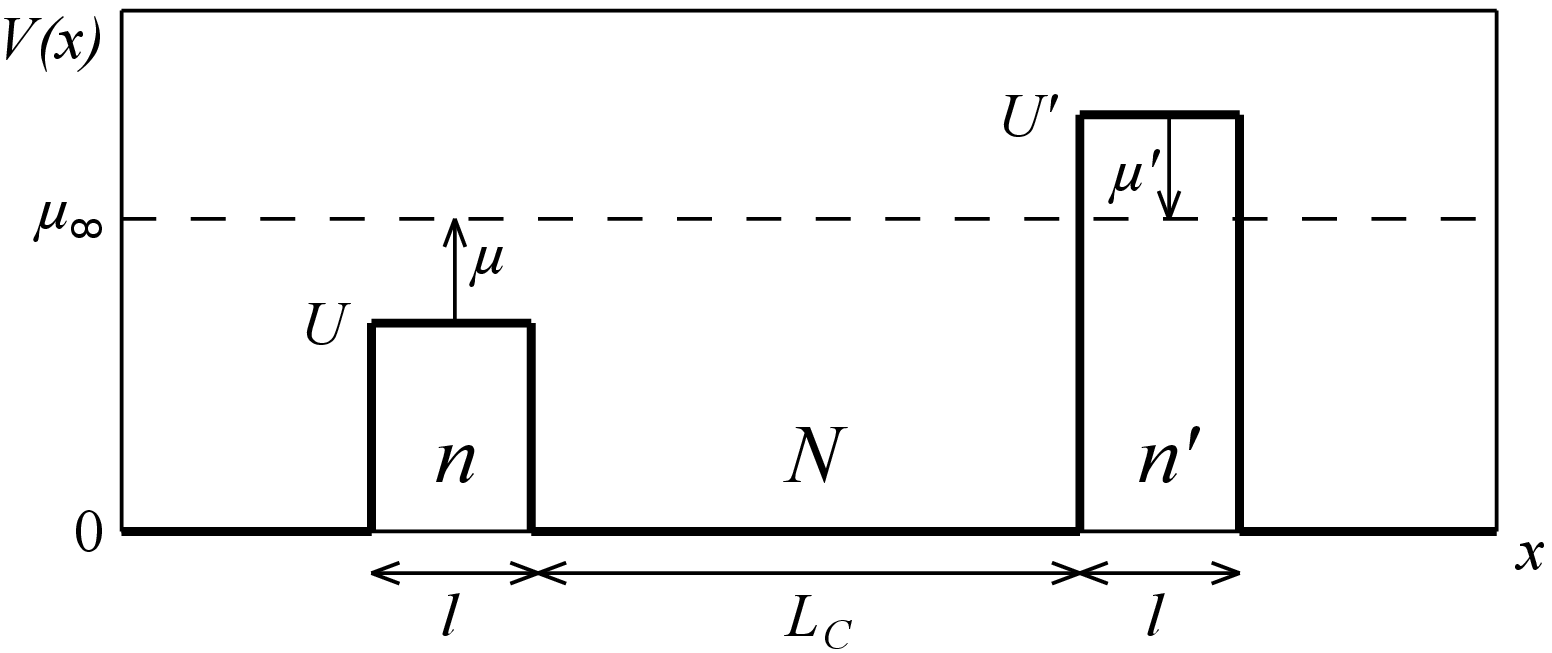}}

\caption{\label{ring}
Schematic diagram of a hexagonal graphene ring attached to graphene leads (top panel) and corresponding potential profile (bottom panel). The constriction with zigzag edges between each lead and the ring allows one to control input and output valley polarizations by varying the electrostatic potentials $U$ and $U'$.
}
\end{figure}

The analysis starts from the tight-binding model of graphene, with Hamiltonian
\begin{equation}
  H=\sum_{i,j}\tau_{ij}|i\rangle\langle j|+\sum_{i}V_{i}|i\rangle\langle i|.
\end{equation}
The system is coupled to the vector potential $\mathbf{A}$ through the hopping matrix element
\begin{equation}
  \tau_{ij}=-\tau\exp\left(\frac{2\pi i}{\Phi_0}\int_{\mathbf{R}_i}^{\mathbf{R}_j}
  d\mathbf{r}\cdot\mathbf{A}\right),
\end{equation}
with $\tau=3\,{\rm eV}$ the hopping energy and $\Phi_{0}=h/e$ the flux quantum. The orbitals $|i\rangle$ and $|j\rangle$ are nearest neighbors on a honeycomb lattice (with lattice points $\mathbf{R}_{i}$), otherwise $\tau_{ij}=0$. The energy-independent velocity $v$ near the Dirac point equals $v=\frac{1}{2}\sqrt{3}\tau a/\hbar\approx 10^6\mbox{ m/s}$, with the lattice constant $a=0.246\mbox{ nm}$.

The electrostatic potential $V_{i}=V(x_{i})$ varies only along the axis connecting the input and output point contacts (see Fig.\ \ref{ring}). Namely, the potential equals $U$ at the first constriction ($0<x<l$, where $l$ is the constriction length), $U'$ at the second constriction ($l+L_C<x<2l+L_C$, with $L_C=(4R-w)/\sqrt{3}$, where $R$ is the ring radius and $w$ is the constriction width), and zero everywhere else. By varying $U$ and $U'$ at a fixed Fermi energy $\mu_\infty$ in the external leads, we can vary the Fermi energies $\mu=\mu_\infty-U$ and $\mu'=\mu_\infty-U'$ in the two constrictions. We took $\mu_\infty=\tau/3$ to work with heavily doped graphene leads, while remaining at sufficiently small Fermi energy that the linearity of the dispersion relation holds reasonably well. The constriction parameters $w=10\sqrt{3}a$ and $l=16a$ are chosen to provide valley polarizations above $90\%$ \cite{Ryc07}. 

We denote the number of transmitted modes through the first constriction by $n$ and through the second constriction by $n'$. A positive number indicates that the Fermi level lies in the conduction band, while a negative number indicates that the Fermi level lies in the valence band. For example, as shown in Ref.\ \cite{Ryc07}, the case $-3\Delta/2<\mu'<0<\mu<3\Delta/2$ (with $\Delta=\pi\hbar v/w$) corresponds to $n=1$, $n'=-1$ (valley filters of opposite polarity), while the case $0<\mu,\mu'<3\Delta/2$ corresponds to $n=n'=1$ (valley filters of the same polarity).

The AB interferometer is modeled by a hexagon with a hexagonal hole in the center and zigzag edges along the entire perimeter. (We will consider a more generic shape in Sec.\ \ref{gensec}.) To vary the number of modes $N$ that can propagate along the ring, we keep the radius fixed at $R=35\sqrt{3}\,a$ and vary the inner radius. We take ring widths $W/\sqrt{3}a=5,10,15$, corresponding to $N=1,3,5$. (Even values of $N$ are not accessible because of the valley degeneracy of the second and higher modes.) The external leads have a fixed width $W_\infty/\sqrt{3}a=70$, corresponding to 29 propagating modes.

We take the vector potential $\mathbf{A}=(A_x,0,0)$ with 
\begin{equation}
\label{axdef}
  A_x=\left\{\begin{array}{cc}
      By, & -\frac{W_\infty-w}{\sqrt{3}} < x < 2l\! +\! L_C\!+\!
      \frac{W_\infty-w}{\sqrt{3}}, \\
      0,  & \mbox{otherwise}. \\
    \end{array}\right. 
\end{equation}
This corresponds to a uniform perpendicular magnetic field $B$ in the area containing the ring, the two point contacts, and the widening region connecting the point contacts to the external leads (see grey rectangle in Fig.\ \ref{ring}, top panel). For technical reasons, the magnetic field is set to zero in the external leads. The resulting step in the field strength at the entrance and exit of the interferometer will reduce its conductance somewhat, but this is not expected to change the qualitative features of the magnetoconductance oscillations that we are interested in.

We calculate the transmission matrix $t$ numerically and then obtain the conductance from the Landauer formula
\begin{equation}
G=\frac{2e^{2}}{h}{\rm Tr}\,tt^{\dagger}.
\end{equation}
(The factor of two accounts for the spin degeneracy.)

\section{Results}
 
\begin{figure}[!p]
\centerline{\includegraphics[width=\linewidth]{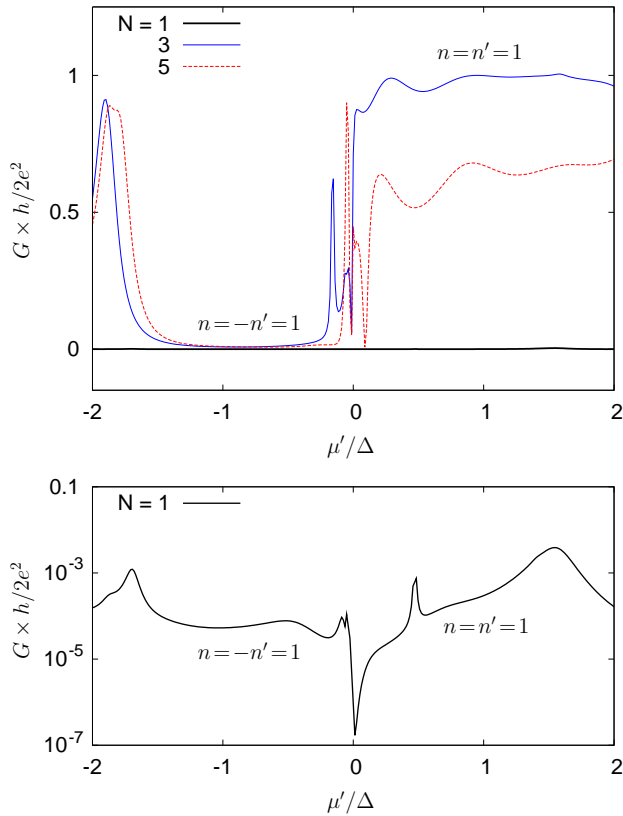}}

\caption{ \label{hexvalve}
Conductance of the hexagonal graphene ring in zero magnetic field at fixed $\mu=0.05\,\tau\approx\Delta/3$ as a~function of $\mu'/\Delta$ (with $\Delta=\pi\tau/20$ for $w=10\sqrt{3}a$). The parameter $\mu'=\mu_{\infty}-U'$ is varied by varying $U'$ at fixed $\mu_{\infty}$. The three curves are for different ring widths, corresponding to $N=1,3,5$ propagating modes along the ring. Bottom panel: The conductance for $N=1$ on an expanded (logarithmic) scale.
}
\end{figure}

\begin{figure}[!p]
\centerline{\includegraphics[width=0.7\linewidth]{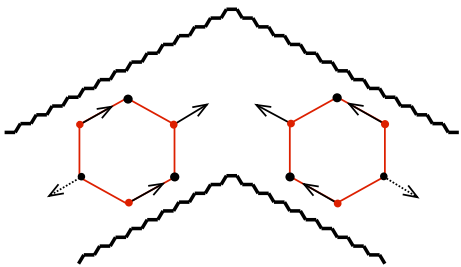}}

\caption{\label{blocking}
Schematic illustration of the mechanism of current blocking by mismatched valley polarizations at a vertex of the hexagonal ring with zigzag edges, responsible for the suppressed conductance when $N=1$. Shown is the first Brillouin zone at the two sides of the vertex, rotated by $\pi/3$. Solid arrows indicate the direction of propagation of the lowest mode in one of the two valleys (located near the three red dots in the Brillouin zone). The lowest mode in the other valley (near the three black dots) propagates in the opposite direction (dotted arrow, shown only near one of the black dots for clarity). The current is blocked because the corresponding points in the Brillouin zone propagate in opposite directions at the two sides of the vertex.
}
\end{figure}

\begin{figure}[!ht]
\centerline{\includegraphics[width=\linewidth]{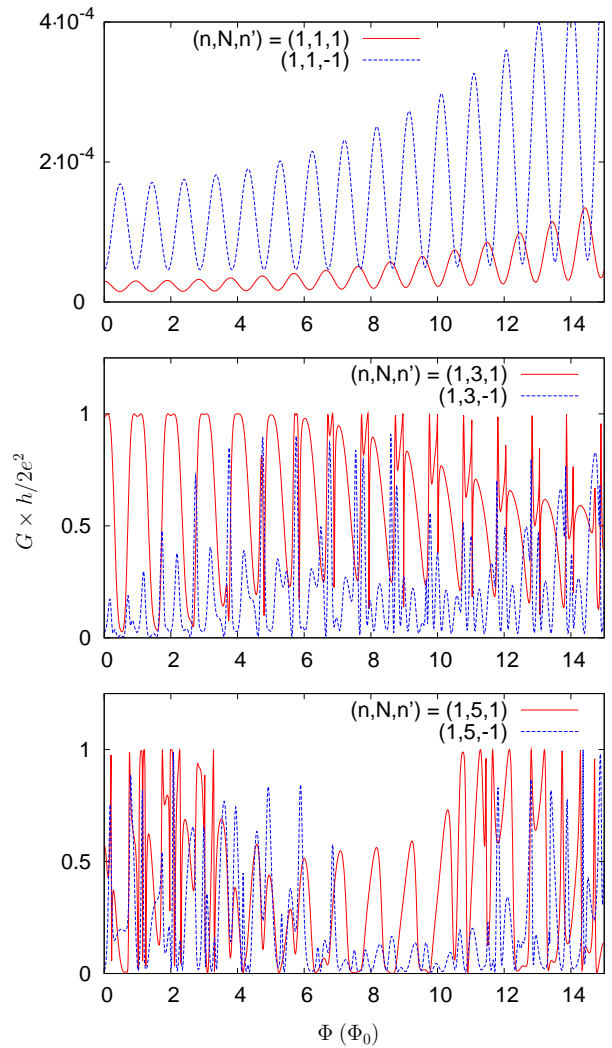}}

\caption{\label{gfi}
Conductance as a function of magnetic flux through the ring. Red solid curves show the case $n=n'=1$, $\mu=\mu'=0.05\,\tau$ of identical valley polarizations in both constrictions, and blue dashed curves show the case $n=-n'=1$, $\mu=-\mu'=0.05\,\tau$ of opposite polarizations. The number $N$ of propagating modes in the ring is varied between the three panels.
}
\end{figure}

\begin{figure}[!ht]
\centerline{\includegraphics[width=\linewidth]{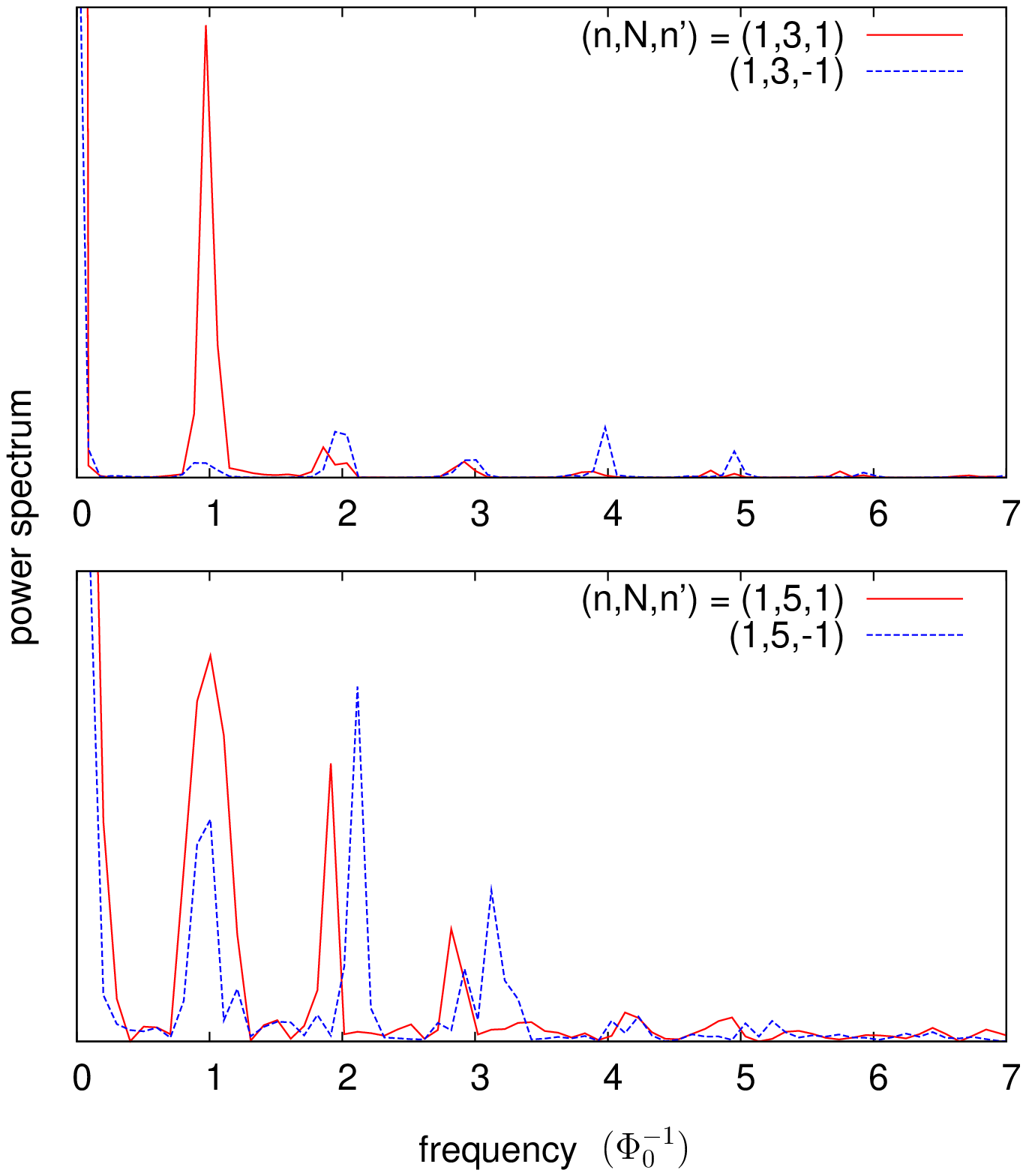}}

\caption{\label{gfft}
Fourier transform of the magnetoconductance data shown in Fig.\ \ref{gfi} for $N=3$ and $5$.
}
\end{figure}

To analyze the operation of the hexagonal graphene ring as a valley valve \cite{Ryc07}, we first show in Fig.\ 2 the conductance in zero magnetic field. We take $\mu=0.05\,\tau$ to keep $n=1$ at the first constriction (the polarizer), and vary the Fermi energy $\mu'$ at the second constriction (the analyzer). For negative $\mu'$ ($n'=-1$), when the two constrictions transmit the opposite valley polarization, we obtain $G\lesssim 10^{-3}\,e^2/h$, showing that intervalley scattering in the system is negligible. The resonances for $\mu'\lesssim -3\Delta/2$ are due to quasi-bound states in the valence band of the second constriction \cite{Sil07}. For positive $\mu'$ ($n'=1$) we would expect a conductance of order $e^2/h$, because the two constrictions transmit the same valley polarization. This is indeed observed for $N>1$, but remarkably enough the conductance remains $\lesssim 10^{-3}\,e^2/h$ if the arms of the ring support a single propagating mode ($N=1$, black line in Fig.\ \ref{hexvalve}). We attribute this anomaly to current blocking from mismatched valley polarizations at the vertices of the hexagonal ring, as explained in Fig.\ \ref{blocking}. 

In Fig.\ \ref{gfi} the conductance $G$ is plotted as a function of the flux $\Phi=(\bar{S}/S_c)\Phi_c$ through the ring, where $\Phi_c$ is the flux per unit cell of area $S_c=\frac{1}{2}\sqrt{3}a^2$, and $\bar{S}=(S_{o}+S_{i})/2$ is the average of the outer and inner ring areas $S_{o}$ and $S_{i}$. The first harmonic frequency of the Fourier spectrum shown in Fig.\ \ref{gfft} is within a few percent of the expected value $\Phi_{0}^{-1}=e/h$, indicating that $\bar{S}$ accurately represents the effective area of the ring.

The harmonic content of the magnetoconductance oscillations is strikingly different when the current is blocked ($N=1$, top panel in Fig.\ \ref{gfi}) and when it is not blocked ($N=3,5$, lower two panels). On the one hand, when the conductance is suppressed below $10^{-3}\,e^{2}/h$ the magnetoconductance oscillations are nearly sinusoidal, almost without higher harmonics. This is as expected for transmission through evanescent modes. On the other hand, when the conductance is of order $e^{2}/h$ the oscillations are highly nonsinusoidal, with appreciable higher harmonics, as expected for transmission through propagating modes.

A feature of the nonsinusoidal magnetoconductance oscillations shown in Fig.\ \ref{gfi}, and quantified by the Fourier transform in Fig.\ \ref{gfft}, is the suppression of the lowest harmonic (period $\Delta\Phi=\Phi_{0}$) in the case of opposite valley polarizations in the two constrictions ($n=-n'=1$). This suppression of the fundamental periodicity is dramatic for $N=3$ (top panels in Figs.\ \ref{gfi} and \ref{gfft}), but it is also noticeable for $N=5$ (lower panels). The second (and higher) harmonics, in contrast, are enhanced in the case of opposite valley polarizations.

The suppression of the first harmonic indicates that electrons which travel only once along the ring have a small probability for intervalley scattering and can therefore not contribute to the conductance when the two constrictions have opposite valley polarization. Higher harmonics correspond to electrons which travel several times along the ring, with a larger probability for intervalley scattering and therefore a larger probability to contribute to the conductance.

\section{Generic graphene ring}
\label{gensec}

\begin{figure}[!ht]
\centerline{\includegraphics[width=\linewidth]{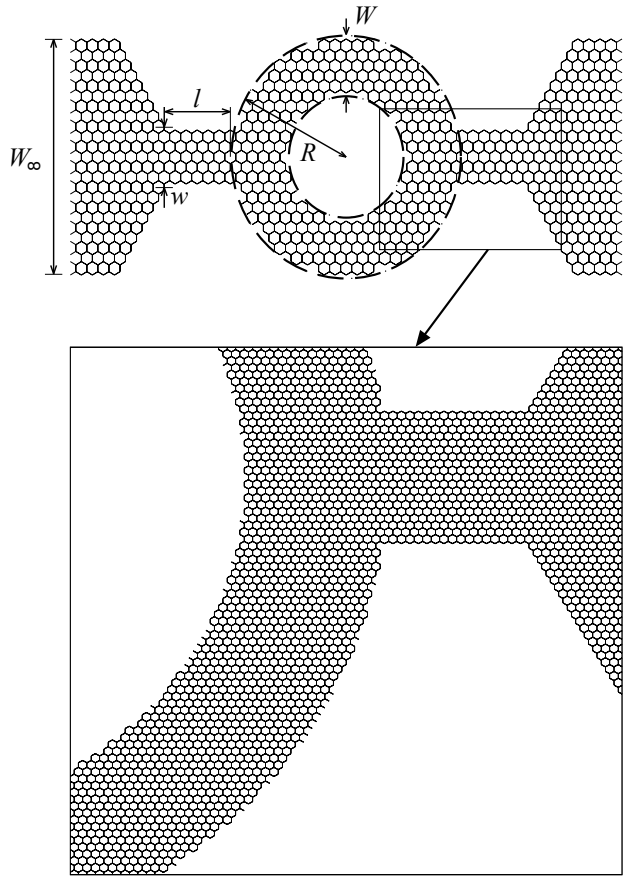}}

\caption{\label{gering}
Graphene ring with an irregular, approximately circular, boundary. Bottom panel: A magnified section of the ring with $R=35\sqrt{3}\,a$ and $W=10\sqrt{3}\,a$ used in our simulations, which shows the irregularity of the boundary.
}
\end{figure}

\begin{figure}[!ht]
\centerline{\includegraphics[width=\linewidth]{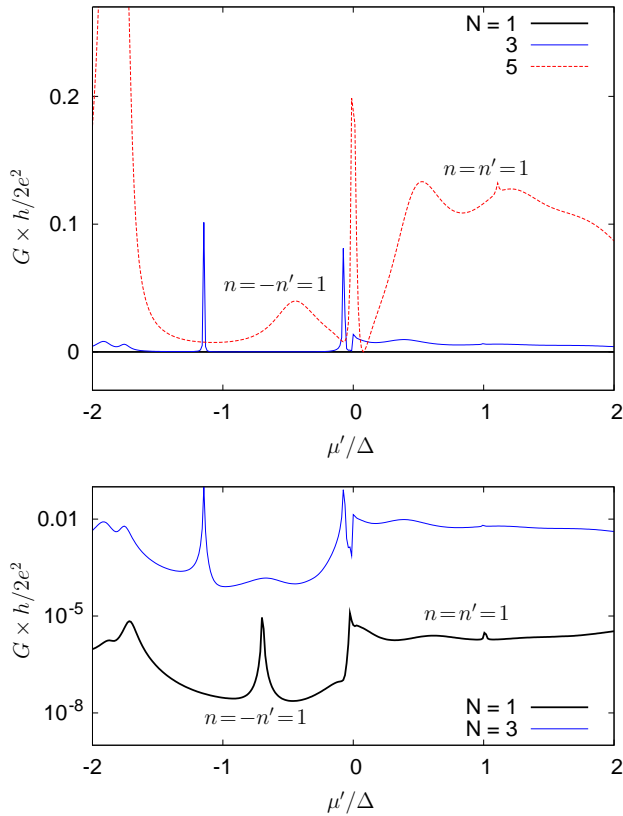}}

\caption{ \label{cirvalve}
Conductance of the circular graphene ring in zero magnetic field at fixed $\mu=0.05\,\tau$ as a~function of $\mu'/\Delta$. The lower panel shows the data for $N=1,3$ on an expanded (logarithmic) scale.
}
\end{figure}

\begin{figure}[!p]
\centerline{\includegraphics[width=\linewidth]{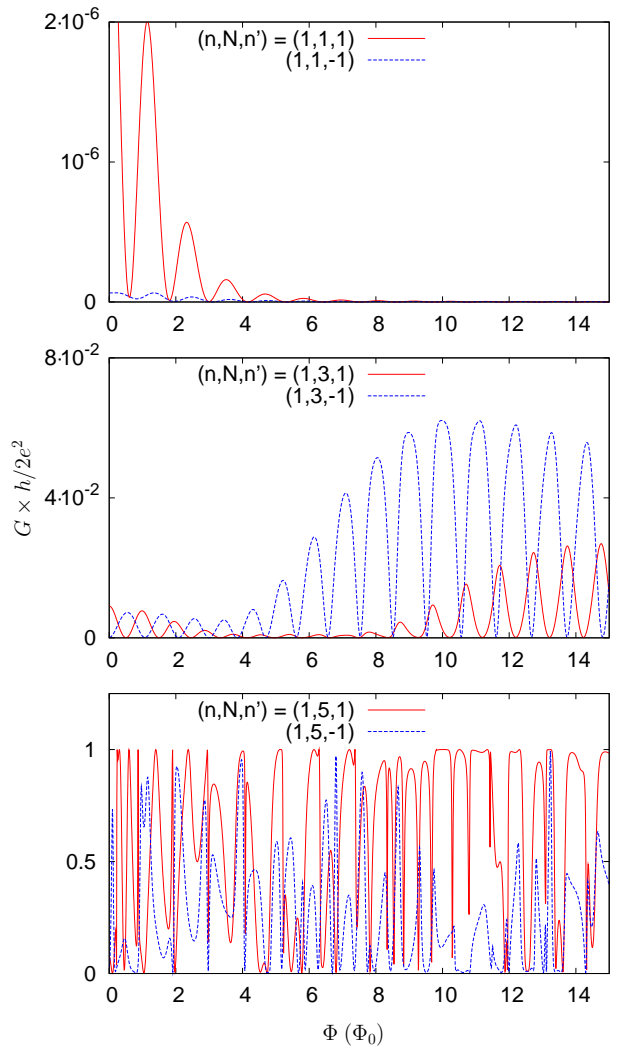}}

\caption{\label{gcifi}
Conductance of the circular ring as a function of magnetic flux.
}
\end{figure}

\begin{figure}[!p]
\centerline{\includegraphics[width=\linewidth]{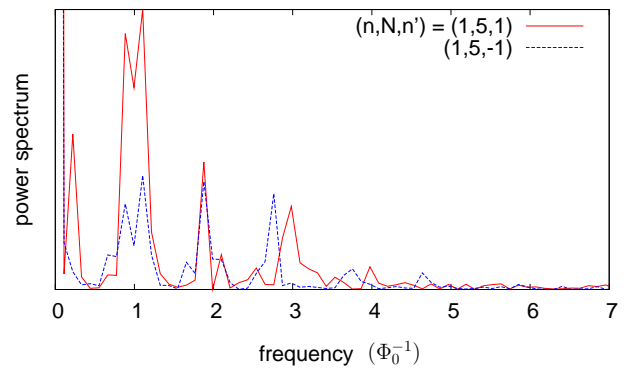}}

\caption{\label{gcifft}
Fourier transform of the magnetoconductance data of Fig.\ \ref{gcifi} for $N=5$.
}
\end{figure}

In order to determine how generic our results for the hexagonal ring might be, we have repeated our calculations for an approximately circular ring with an irregular boundary (see Fig.\ \ref{gering}). We constructed the circular ring by starting from the hexagonal ring and then keeping only lattice sites within an annulus formed by two concentric circles. The electrostatic potential shown in the bottom panel of Fig.\ \ref{ring} and the vector potential given by Eq.\ (\ref{axdef}) remain unchanged, only the length of the central area $L_C$ is replaced now by $L_C'=\sqrt{4R^2-w^2}$. The point contacts still have a zigzag boundary, so the filtering property at entrance and exit should remain intact, but the circular ring no longer has zigzag boundaries along its entire perimeter (as it did in the case of the hexagonal ring). Results are shown in Figs.\ \ref{cirvalve}--\ref{gcifft}.

Because the boundaries of the ring are not uniform, the number of propagating modes in the ring is not well-defined. We still label the data by the same number $N$ as in the hexagonal ring, but now this number refers only to the number of modes which the ring would support if the boundaries were of zigzag form. This uncertainty in the definition of $N$ might explain why we now see the current blocking not only for $N=1$ but also for $N=3$ (Fig.\ \ref{cirvalve}, lower panel). Only for $N=5$ do we obtain an appreciable current through the ring. The conductance for $N=5$ is still well below the optimal value of $2e^{2}/h$, but more than $100$ times larger than for $N=1,3$.

The harmonic content of the magnetoconductance oscillations (Figs.\ \ref{gcifi}, \ref{gcifft}) is qualitatively similar as in the case of the hexagonal ring. When the current is blocked ($N=1,3$) we see nearly sinusoidal oscillations, almost without higher harmonics. When the current is not blocked ($N=5$) higher harmonics do appear and the lowest harmonic is strongly suppressed when the polarity of the valley filters in the two point contacts is opposite.

\section{Conclusions}

In conclusion, we have identified signatures of valley polarization in the magnetoconductance of an Aharonov-Bohm interferometer in graphene. The suppression of the lowest harmonic of the conductance oscillations that appears when the two point contacts have opposite valley polarity indicates that electrons which have travelled only once along the ring preserve their valley polarization, and therefore cannot contribute to the conductance. 

The special case of a single-mode hexagonal ring shows nearly sinusoidal magnetoconductance oscillations around a greatly suppressed average conductance, attributed to a current blocking effect at a vertex where two zigzag edges meet at an angle of $\pi/3$. This current blocking is not a special feature of the hexagonal geometry. Instead, we have found the same current blocking in a circular ring with irregularly shaped boundaries. This finding is consistent with a recent theory for the boundary condition of a terminated honeycomb lattice at an arbitrary crystallographic orientation \cite{Akhmerov:unpublished}. It is found that the zigzag boundary condition applies generically for any angle $\phi\neq 0\;({\rm mod}\;\pi/3)$ of the boundary (where $\phi=0$ labels the armchair orientation). The current blocking mechanism illustrated in Fig.\ \ref{blocking} for the vertex between two zigzag boundaries would then apply generically to any boundary near an orientation of $\phi=0\;({\rm mod}\;\pi/3)$.

\section*{Acknowledgment}
This work was supported by the Dutch Science Foundation NWO/FOM. Discussions with A.R. Akhmerov and P. Recher are gratefully acknowledged. One of the authors (A.R.) acknowledges the support from a special grant by the Polish Science Foundation (FNP) and by the Polish Ministry of Science (Grant No. 1--P03B--001--29).

\end{document}